\documentclass{aa}

\usepackage{amsmath}	
\usepackage{orcidlink}
\usepackage{graphicx}
\usepackage{xcolor}
\usepackage{txfonts}
\usepackage{natbib}
\usepackage{hyperref}
\hypersetup{
    colorlinks=true,
    citecolor=blue,
    linkcolor=blue,
    urlcolor=blue,}
\usepackage{longtable}
\usepackage{caption}

\makeatletter
\renewcommand*\aa@pageof{, page \thepage{} of \pageref*{LastPage}}
\makeatother

\bibpunct{(}{)}{;}{a}{}{,}

\newcommand{\pr}{\ensuremath{P_{\rm r}}}

\newcommand{\nullc}{\multicolumn{2}{c}{}}
\newcommand{\pmax}{\ensuremath{P_{\rm max}}}
\newcommand{\pmin}{\ensuremath{P_{\rm min}}}
\newcommand{\amin}{\ensuremath{\alpha_{\rm min}}}
\newcommand{\amax}{\ensuremath{\alpha_{\rm max}}}
\newcommand{\ainv}{\ensuremath{\alpha_0}}
\newcommand{\FAtt}{FA22}

\begin{document}
\title{Polarimetry and albedo of the Near-Earth Asteroid 2025 FA22}
\titlerunning{Polarimetry of NEA 2025 FA22}

\author{
J.-P.~Rivet\inst{1}{\orcidlink{0000-0002-0289-5851}}       \and
S.~Bagnulo\inst{2}{\orcidlink{0000-0002-7156-8029}}        \and
P. Bendjoya\inst{1}{\orcidlink{0000-0002-4278-1437}}       \and
G.~Borisov\inst{3,2}{\orcidlink{0000-0002-4516-459X}}      \and
A.~Cellino\inst{4}{\orcidlink{0000-0002-6645-334X}}        \and
M.~Devog\`ele\inst{5}{\orcidlink{0000-0002-6509-6360}}     \and
Z.~Gray\inst{6}{\orcidlink{0000-0002-6610-1897}}           \and
S.~Ieva\inst{7}{\orcidlink{0000-0001-8694-9038}}           \and
L.~Kolokolova\inst{8}{\orcidlink{0000-0002-9321-3202}}     \and
Y.~G.~Kwon\inst{9}{\orcidlink{0000-0002-8122-3606}}        \and
A. Berdyugin\inst{10}{\orcidlink{0000-0002-9353-5164}}     \and
S.~V.~Berdyugina\inst{11,12}{\orcidlink{0000-0002-2238-7416}} \and
L.~Boulanger\inst{1}{\orcidlink{0000-0002-4127-2649}}      \and
E.~Dotto\inst{7}{\orcidlink{0000-0002-9335-1656}}          \and
P.~Fatka\inst{13}{\orcidlink{0009-0007-4890-1667}}         \and
E.~Frank\inst{2}{\orcidlink{0000-0001-6436-8769}}          \and
M.~Lazzarin\inst{14}{\orcidlink{0000-0001-7976-2339}}      \and
V.~Piirola\inst{9}{\orcidlink{0000-0003-0186-206X}}        \and
P.~Pravec\inst{13}{\orcidlink{0009-0007-4890-1667}}         \and
the NEOPOPs team
}

\institute{
Observatoire de la C\^{o}te d'Azur, Universit\'{e} C\^{o}te d'Azur, CNRS
Bd de l'Observatoire, CS~34229, 06304 Nice Cedex 4, France
\and
Armagh Observatory \& Planetarium, College Hill, Armagh BT61 9DG, UK.
\and
Institute of Astronomy and National Astronomical Observatory, Bulgarian Academy of Sciences, 72 Tsarigradsko Chauss\'ee Blvd., BG-1784 Sofia, Bulgaria. \email{gborisov@nao-rozhen.org}
\and
INAF -- Osservatorio Astrofisico di Torino, via Osservatorio 20, I-10025 Pino Torinese, Italy
\and
ESA NEO Coordination Centre, European Space Agency, Largo
Galileo Galilei, 1, Frascati, 00044, RM, Italy
\and
Department of Physics, University of Helsinki, PO Box 64, 00014, Finland
\and
INAF Roma
\and
Department of Astronomy, University of Maryland, College Park, MD 20742-2421, US
\and
Caltech/IPAC, 1200 E California Blvd, MC 100-22, Pasadena, CA 91125, USA
\and
Department of Physics and Astronomy, FI-20014 University of Turku, Finland
\and
Istituto Ricerche Solari Aldo e Cele Dacc\`o (IRSOL), Faculty of Informatics, Universit\`a della Svizzera italiana, Via Patocchi 57, Locarno, Switzerland
\and
Euler Institute, Faculty of Informatics, Universit\`a della Svizzera italiana, Via Buffi 13, 6900 Lugano, Switzerland
\and
Astronomical Institute of the Academy of Sciences of the Czech Republic, Fri\v{c}ova 298, Ond\v{r}ejov, CZ-25165, Czech Republic
\and
INAF Padova
}

   \date{Received ; accepted }

\abstract{
We report spectropolarimetric and broadband polarimetric observations of the near-Earth asteroid 2025~FA22 during its close approach of 18~September 2025 (about two Moon distances). With a diameter estimated between $130$ and $290$~m, 2025 FA22
is among the largest NEAs observable at such proximity, prompting an International Asteroid Warning Network (IAWN) rapid-response campaign. Although early orbital solutions indicated a possible impact in 2089, further follow-up astrometric observations ruled out collision hazard. The favourable geometry of this close encounter enabled a dense coverage of the positive part of the phase-polarisation curve, from the high polarisation domain (high phase angles), nearly to the inversion angle where the linear polarisation fraction vanishes. The spectropolarimetric observations provided the wavelength dependence of the polarisation fraction. Using empirical relationships, an estimate of the geometric albedo could be drawn from the slope of the phase-polarisation curve at inversion angle\,: $\rho_v=0.17\pm0.04$ in the V~band. This value, together with the spectropolarimetric trend, provides constraints on the taxonomic class, with the results being most consistent with an M (or Xc)-type classification. These results demonstrate the interest of polarimetry and spectropolarimetry for rapid characterisation of newly discovered NEAs in planetary defence campaigns.
}

\maketitle

%

\section{Introduction}\label{Sect_Intro}
On 18 September 2025, the near-Earth asteroid (NEA) 2025 FA22 (hereafter \FAtt) had a close approach to Earth, reaching a minimum distance of about two lunar distances. With an estimated diameter between 130 and 290\,m, \FAtt\ is among the largest NEAs observed at such close range. Initially flagged as a possible impactor in 2089, and briefly topping ESA’s Risk List\footnote{\url{https://neo.ssa.esa.int/risk-list}}, subsequent astrometric follow-up quickly refined its orbit and excluded any imminent impact scenario. Although \FAtt\ poses no imminent hazard, its encounter presented a rare opportunity to carry out coordinated, multi-technique observations in the framework of the International Asteroid Warning Network (IAWN), a United Nations–endorsed consortium that coordinates global observations, data sharing, and communication on potential impact threats. In particular, the \FAtt\ campaign was aimed at testing the planetary defence community’s readiness to rapidly characterise a newly discovered NEA.

In this framework, polarimetry is particularly powerful. By measuring the linear polarisation of scattered sunlight as a function of phase angle (the angle between the target--Sun and target--observer directions), it is possible to derive key surface properties including albedo, regolith structure, and, when spectropolarimetric data are collected, also taxonomic classification (see \citealt{Bagnulo2024} for a review).

Main-belt asteroids have been extensively studied with polarimetry, but geometrical constraints limit their ground-based observations to phase angles below $30-40^\circ$, where polarisation is small ($\la 3$\,\%). Near-Earth asteroids, in contrast, can be observed at much larger phase angles where the fraction of linear polarisation becomes much higher, providing a golden opportunity to probe a mostly unexplored regime of the asteroid polarimetric behaviour, including a better understanding of the polarisation–albedo relationship.
Such measurements are directly relevant for planetary defence: coupled with brightness measurements, albedo is a key parameter to obtain reliable estimates of an object's size. Moreover, surface properties inferred from polarimetric data influence the design and execution of any possible deflection attempt whenever this turns out to be needed.

\begin{figure*}[htb]
\centering
\includegraphics[width=0.9\textwidth]{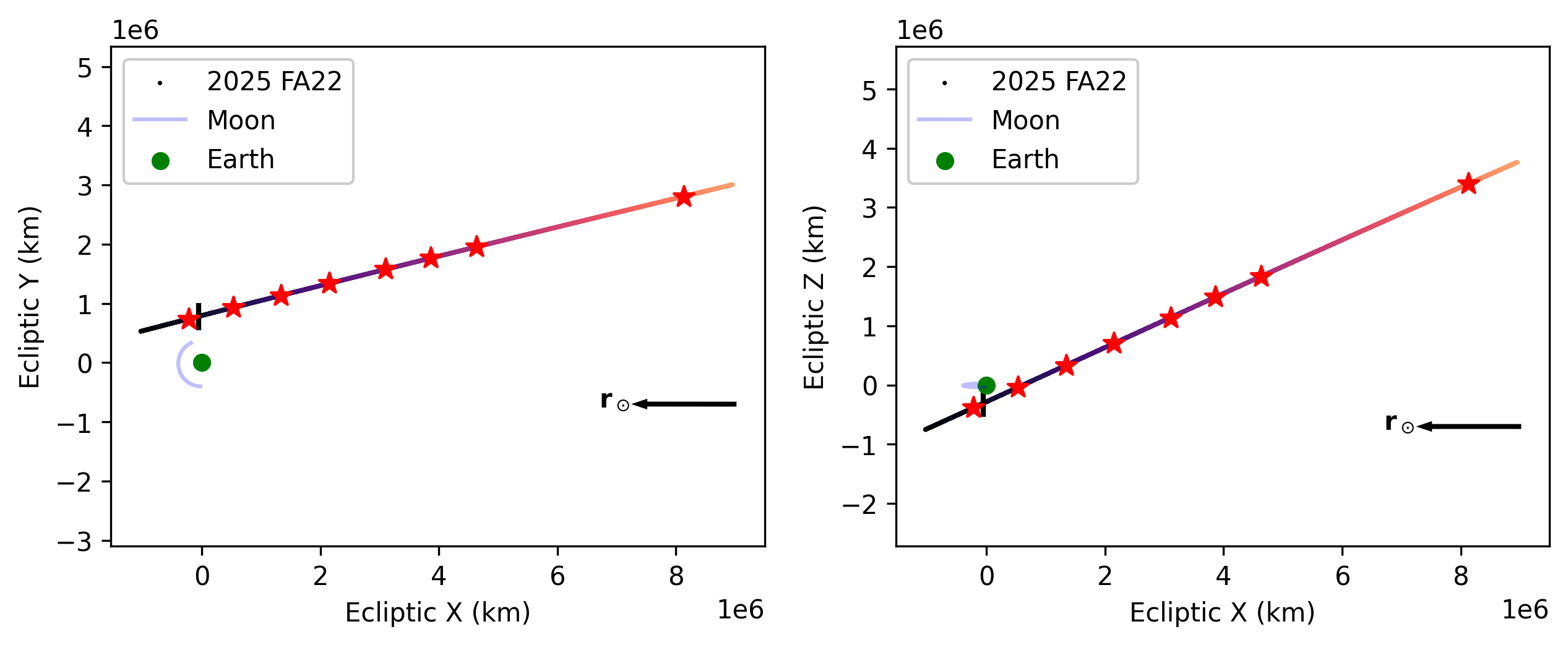}
\caption{Observing geometry of \FAtt\ relative to the Moon and Earth in the projected ecliptic $x$--$y$ (left) and $x$--$z$ (right) planes. Star symbols mark the UTs of the first observation on each date listed in Table~\ref{tab:ObservingLog}. The vertical bar indicates the projected point at closest Earth approach (UT 2025-09-18T07:39:48.108+00:00; distance $\approx$842,060\,km). The solid black arrow indicates the anti-solar vector at the median epoch of the observing campaign.}
\label{fig:obs_geo}
\end{figure*}

Here we report spectropolarimetric and multi-band filter-polarimetric measurements of \FAtt\ obtained during its close approach. Our observing strategy exploited the brightness evolution of the object: spectropolarimetry was carried out when \FAtt\ was near its maximum brightness, enabling reliable measurement of the wavelength dependence of polarisation. Broadband filter polarimetry was used when the object was fainter, in order to efficiently sample the phase-polarisation curve with minimal telescope time.

Figure \ref{fig:obs_geo} shows the observing geometry of \FAtt\ relative to the Earth and Moon in the projected ecliptic
$x$–$y$ (left) and $x$–$z$ (right) planes during the campaign. The first date of each observing epoch is marked along the asteroid’s path. During the campaign, the NEA passed close to the Earth–Moon system, reaching a minimum geocentric distance of about 842,060\,km.

For atmosphere-less bodies in the solar system, the polarisation state of the diffused solar light is estimated through the ``linear polarisation fraction'' defined as\,:
\begin{equation}
\pr (\alpha,\lambda)= \frac{F_\perp - F_\parallel}
                    {F_\perp + F_\parallel}
\label{Eq_Pr}
\end{equation}
where $F_\parallel$ and $F_\perp$ are the fluxes of the light polarised in the direction parallel or perpendicular to the scattering plane, respectively. 
A positive sign for \pr\ means that the light flux polarized perpendicularly to the scattering plane exceeds the flux polarized parallel to it, whereas a negative sign means that the parallel polarization dominates.
The linear polarisation fraction (hereafter called by the shortcut ``polarisation'') depends on the wavelength and on the phase angle.
The phase–polarisation curve shows two distinct branches. At small phase angles, typically $\la 20^\circ$, the flux polarized parallel to the scattering plane dominates the flux polarized perpendicularly; this is called the ``negative branch''. At larger phase angles however, the flux polarized perpendicularly dominates, forming the ``positive branch''. In our data set, we achieved dense coverage of the positive branch, starting just above the inversion angle \ainv\ (that separates the two branches) and extending up to the phase angles where the polarisation reaches its maximum. Here we present our data, and we use them to estimate the albedo of the object, via semi-empirical formulas that link the albedo to the characteristics of the phase-polarisation curve (e.g., \citealt{Cellino2012,Cellino2016}). Additional constraints from spectropolarimetry also allow us to set some constraints on the spectral class of the asteroid.

\section{Observations}
We have measured the linear polarisation fraction as a function of wavelength $\lambda$ and phase angle $\alpha$. In the case of filter polarimetry, $\lambda$ refers to the effective wavelength of the filter employed in the observations. 

Observations were obtained between 2025-09-18 and 2025-09-28 using both spectropolarimetry and filter polarimetry, covering phase angles from $109^\circ$ to $24.2^\circ$ (see the full observing log in Table~\ref{tab:ObservingLog}). Data were collected at multiple observatories with different instruments, all employing the beam-swapping technique \citep[e.g.,][]{Bagetal09}.

Filter polarimetry was performed with nominally similar B, V, and R filters whose transmission curves differed slightly between instruments. 
Table~\ref{tab:Filters} summarizes the characteristics of the filters involved in the filter polarimetry measurements. In this table, $\lambda_0$ is the central wavelength, and $\Delta\lambda$ is the FWHM. For DIPol-UF, the given values combine the effect of the filters and of the dichroic beam splitters, as discussed in \citet{BerPii22}.
    \begin{table}[h]
        \centering
        \caption{Filters definition for filter polarimetry measurements.}
        \label{tab:Filters}
        \begin{tabular}{|l|l|l|l|}
             \hline
             Instrument & Filter     & $\lambda_0$ (nm) & $\Delta\lambda$ (nm) \\
             \hline
             FoReRo2    & R          & $680$            & $111$                \\
             DIPol-UF   & R'         & $640$            & $96$                 \\
             DIPol-UF   & V'         & $540$            & $75$                 \\
             DIPol-UF   & B'         & $445$            & $114$                \\
             ALFOSC     & R\_Bes     & $650$            & $130$                \\
             ALFOSC     & B\_Bes     & $440$            & $100$                \\
             FORS2      & R\_SPECIAL & $655$            & $165$                \\
             FORS2      & V\_HIGH    & $555$            & $123$                \\
             \hline
        \end{tabular}
    \end{table}

While such differences can be critical in cases where polarisation varies rapidly with wavelength, asteroid polarisation varies smoothly across the optical range. As a result, small differences in filter transmission have a negligible effect compared with photon-noise uncertainties for this target. This assumption was confirmed a posteriori through numerical simulations based on FORS spectropolarimetry combined with different filter transmission functions, and by the agreement, within uncertainties, of observations obtained at nearly identical phase angles with different instruments. Although detailed information on the exact filters is provided in the relevant instrument documentation (see below), we will hereafter refer to the data simply as B, V, and R band observations.

\subsection{Filter polarimetry with DIPol-UF}
Multicolour filter polarimetry of \FAtt\ was obtained with the Double Image Polarimeter Ultra Fast (DIPol-UF) described by \citet{Piirola2021}.
This three-channel (B', V', R') polarimeter is attached to the f/12.5 Cassegrain focus of the Omicron@C2PU telescope (diameter 1.03\,m), located at the Calern observing station of Observatoire de la C\^ote d'Azur, France. The setup involves a super-achromatic half-wave plate (HWP) as the modulator and a plane-parallel calcite plate as the polarising beam splitter. The polarimeter simultaneously records two orthogonally polarised beams from each source in the field of view. Dichroic beam splitters allow for simultaneous recording on three dedicated CCD cameras in B', V' and R' filters (see Table~\ref{tab:Filters} and a more complete description in \citet{BerPii22}). Each cycle involved 16 positions of the retarder plate, and for each position, three images with 90~s exposure time were taken simultaneously by the B', V' and R' cameras. The telescope was in differential tracking mode (following the proper motion of the asteroid). The target's high proper motion combined with long exposure times yielded long trailed images for the surrounding stars. Thus, some images had to be rejected because of stellar contamination. Data were reduced following the pipeline described by \citet{Piirola2021} and references therein. DIPol-UF observations spanned from 2025-09-18 03:42 UT (phase angle $107.47^\circ$) to 2025-09-23 22:02 UT (phase angle $28.58^\circ$). See table~\ref{tab:ObservingLog} for more details.

\subsection{Spectropolarimetry with FORS2}
\label{Sect_SPol}
We obtained spectropolarimetric observations of \FAtt\ using the FORS2 instrument \citep{appetal98} on the ESO Very Large Telescope (VLT). FORS2 is a slit-fed spectrograph equipped with a retarder waveplate and a Wollaston prism acting as a beam-splitter polariser. Observations were performed with the 300V grism without using any order-sorting filter, providing spectral coverage between 370 and 930\,nm. (\citeauthor{BagNazeetal17} \citeyear{BagNazeetal17} have demonstrated that FORS2 spectropolarimetry with the 300V grism is not affected by order contamination). Slit width was set to 2\arcsec, for a spectral resolving power $R\simeq 200$. These spectropolarimetric observations spanned from 2025-09-18 08:16 UT (phase angle $96.98^\circ$) to 2025-09-20 05:28 UT (phase angle $41.85^\circ$). See table~\ref{tab:ObservingLog} for more details. Data were reduced using the ESO FORS pipeline \citep{Izzetal10}, supplemented by dedicated Fortran routines. Two frames obtained on 2025-10-18 were discarded in our analysis, due to background star contamination; the polarisation spectra were inconsistent with those obtained in other epochs. 

Spectropolarimetric observations of main-belt asteroids of Fig.~\ref{Fig_PolSpectrum} were also obtained with the same configuration. The spectrum of asteroid (433) Eros was obtained on 2014-02-15 at $\alpha=42.0^\circ$, and already published by \citet{Bagetal15}; that of (45) Eugenia was obtained on 2015-06-18 at $\alpha = 21.8^\circ$, and that of (21) Lutetia was obtained on 2015-06-02 at $\alpha=27.5^\circ$.

\subsection{Filter polarimetry with FORS2}
In addition to spectropolarimetry, the FORS2 instrument on the ESO VLT was also used to obtain broadband polarimetric measurements, using the R\_special and v\_high filters (see Table~\ref{tab:Filters}). These filter polarimetric observations spanned from 2025-09-22 03:40 UT (phase angle $32.20^\circ$) to 2025-09-28 04:05 UT (phase angle $24.17^\circ$). See table~\ref{tab:ObservingLog} for more details. The instrument position angle was aligned to the direction perpendicular to the scattering plane. We measured only the reduced $Q/I$ parameter, setting the retarder waveplate to position angles 0, 45, 90, $135^\circ$; observations at the intermediate angles 22.5, 67.5, 112.5 and $157.5^\circ$ were skipped, as in a reference system aligned with the scattering plane, the $U/I$ parameter is zero for symmetry reasons. Data were reduced as described by \citet{Bagetal16} and \citet{Bagetal23}. 

\subsection{Filter polarimetry with FoReRo2}
We also obtained imaging polarimetry in the R filter (see Table~\ref{tab:Filters}) with the two-channel Focal Reducer Rozhen (FoReRo2) \citep{FoReRo2-Jockers,NikolovEtAl2026} attached to the 2-m Ritchey–Chr\'etien-Coudé (2mRCC) telescope of the Bulgarian National Astronomical Observatory (BNAO) Rozhen during four nights from 2025-09-18 02:49 UT (phase angle $109.43^\circ$) to 2025-09-20 21:33 UT (phase angle $36.92^\circ$). See table~\ref{tab:ObservingLog} for more details.
The observations were carried out for 8 retarder waveplate angles. The field of view was generally very crowded, and sometimes the target overlapped the background star trails. For each retarder position angle, we retained three frames in which the asteroid image was not contaminated by star trails.

\subsection{Filter polarimetry with ALFOSC}
We obtained imaging polarimetric observations in the R\_Bes and B\_Bes filters (see Table~\ref{tab:Filters}) with the Alhambra Faint Object Spectrograph and Camera (ALFOSC) attached to the F/11 focus of the $2.56$~m Nordic Optical Telescope (NOT) during three nights from 2025-09-18 05:09 UT (phase angle $104.24^\circ$) to 2025-09-24 04:14 UT (phase angle $28.19^\circ$). See table~\ref{tab:ObservingLog} for more details. The polarimetric setup consists of a rotatable retarder waveplate followed by a calcite plate as a beam-splitting device. Observations in each filter consisted of exposures at 16 different retarder waveplate angles. As in the case of FoReRo2, a number of frames were removed due to background star contamination. Data were reduced using dedicated Python scripts. Further description of the instrument and data reduction technique are outlined in \citet{Grayetal2024}.

\section{Results}
The observing log and the polarisation measurements are given in Table~\ref{tab:ObservingLog}.
Polarimetric data were collected in the B, V, and R bands. The FORS2 spectropolarimetric values of Table~\ref{tab:ObservingLog} were convolved with the transmission curves of the DIPol-UF filters \citep[see, e.g., Eqs.~(19) of][]{BagCoxetal17}.

We have also determined the polarimetric wavelength gradient, or polarimetric colour, defined as\,:
\begin{equation}
PC(\lambda_1, \lambda_2) = \frac{\pr(\lambda_2) - \pr(\lambda_1) }{(\lambda_2 - \lambda_1)}~,
\end{equation}
\noindent where $\lambda_2 > \lambda_1$ are the central wavelengths of two consecutive bands (R and V, or V and B). Table~\ref{tab:Filters} summarizes the central wavelengths of the spectral bands of this work. The polarimetric colours $PC$(B,V) and $PC$(V,R) reported in Table~\ref{tab:ObservingLog} refer to the effective wavelengths of the R, V, B filters, and are expressed in \% per nm.

In the following, we analyse the behaviour as a function of wavelength $\lambda$ and phase angle $\alpha$ of the quantity $\pr(\alpha,\lambda)$.

\subsection{The phase angle polarisation curve and albedo determination}
\begin{figure*}
\centerline{\includegraphics[angle=0,width=16cm,clip]{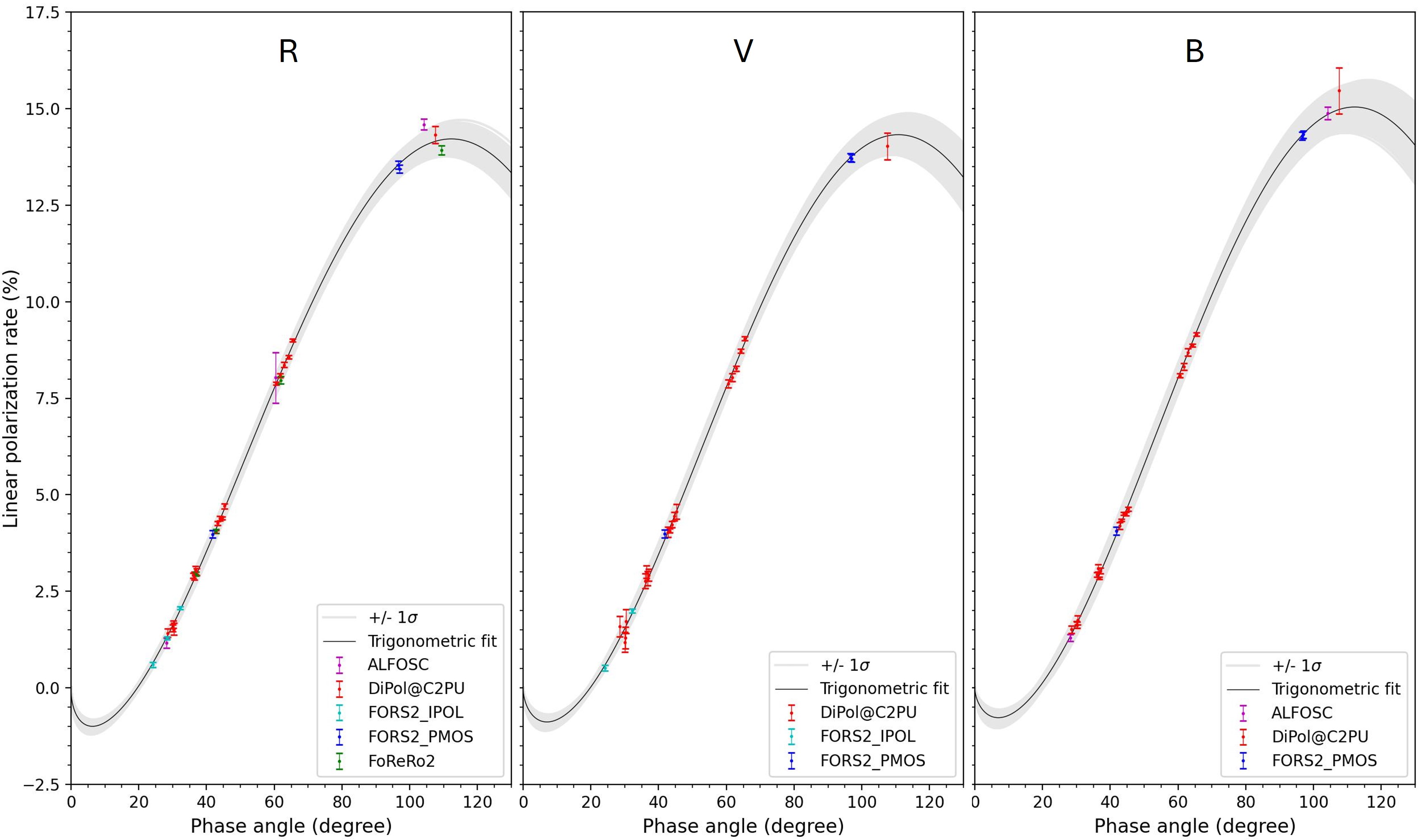}}
\caption{Phase-polarisation curves in R~band (left), V~band (centre), and B~band (right). The solid lines correspond to the best-fits obtained with Eq.~(\ref{Eqn:trigo}) and the grey zones to the $\pm1\,\sigma$ boundaries.}
\label{Fig_PolCurve}
\end{figure*}

Figure~\ref{Fig_PolCurve} shows the \pr values as a function of the phase-angle. The morphology of these phase-polarisation curves can be characterised by parameters which are linked with important physical characteristics of the object, including the geometric albedo.
As mentioned above, because NEAs can be observed at small distances from the Earth, observations can cover much wider intervals of phase angle, as compared with main-belt asteroids.

The behaviour is characterised by a moderately steep polarimetric slope. At phase angles around $50^\circ$, \pr\ reaches values slightly larger than 5\% in all filters, suggesting already an albedo value significantly higher than in the case of some low-albedo NEAs, such as (3200) Phaethon \citep{Devetal18}.

\begin{figure}
\begin{center}
\includegraphics[angle=0,width=8.8cm,trim={0.8cm 6.0cm 1.0cm 2.5cm},clip]{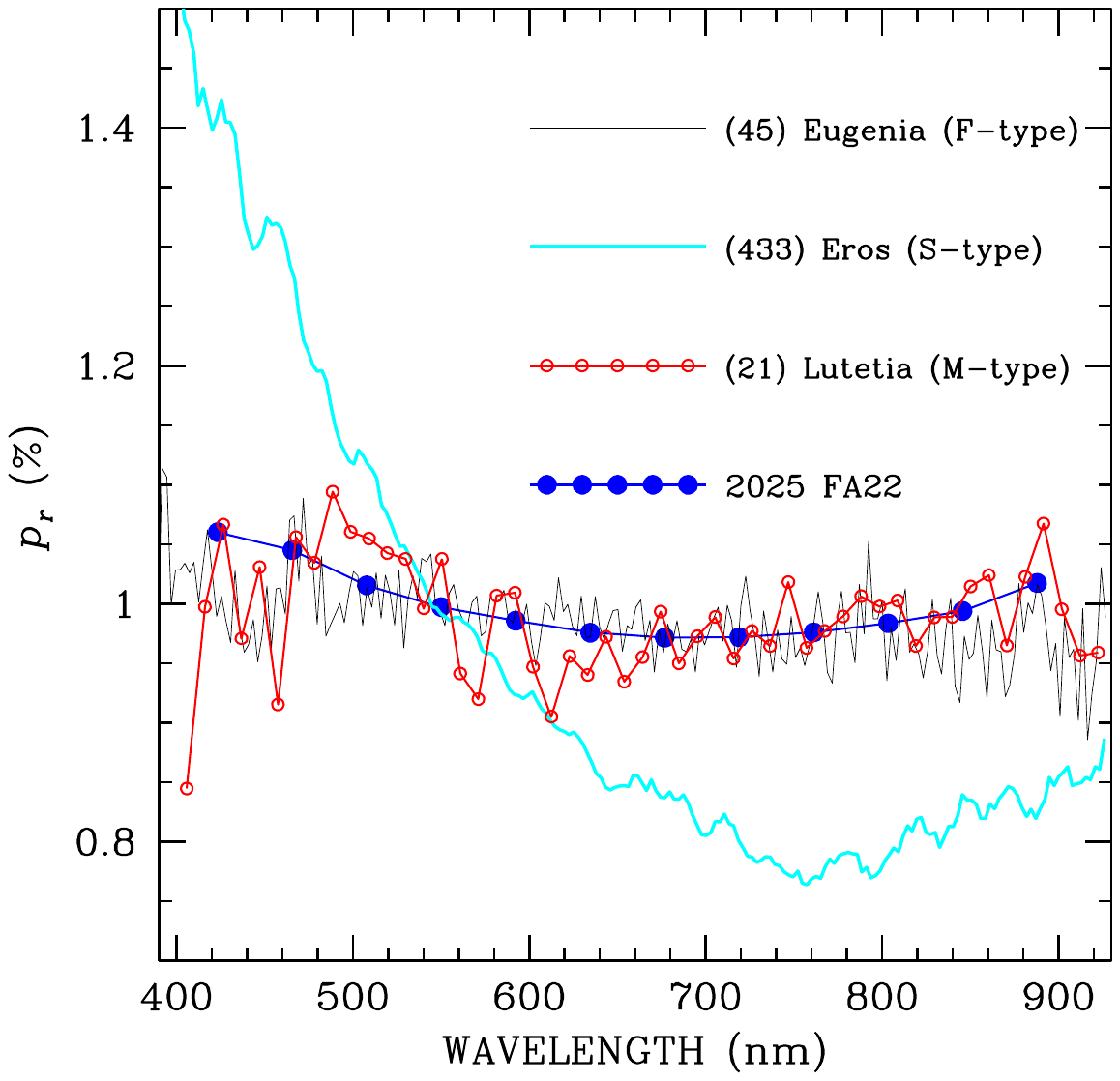}
\caption{Normalised polarisation spectrum of \FAtt\ obtained on 2025-09-18 at $\alpha \simeq 97^\circ$ (blue solid circles)
compared with the spectra of other asteroids (obtained in the positive branch).}
\label{Fig_PolSpectrum}
\end{center}
\end{figure}

We computed a best-fit of the data shown in Fig.~\ref{Fig_PolCurve} by using the function
\begin{equation}
    \pr(\alpha) = A\, \sin^B(\alpha) \, \cos^C\left(\frac{\alpha}{2}\right) \, \sin(\alpha - \ainv)
\label{Eqn:trigo}
\end{equation}
where $A$, $B$,  $C$ and the inversion phase angle \ainv\  are parameters to be determined by means of best-fit techniques. 
Equation~(\ref{Eqn:trigo}) is the so-called trigonometric representation of the
phase-polarisation relationship of asteroids originally proposed by \citet{LumMuin93}, and later adopted by other authors, including, for instance, \citet{Pentetal05} and \citet{Devetal18}. This representation of the phase-polarisation curve is normally used to fit
NEA polarimetric data, because it provides a satisfactory
representation of observations obtained up to high phase angles, well beyond the limits reached by main-belt asteroids.

\begin{table*}
\caption{\label{tab:Coeffs} Best-fit coefficients and characteristic parameters of the polarimetric curve, with their $1\,\sigma$ uncertainties.}
\tabcolsep=0.14cm
\begin{center}
  \begin{tabular}{c r@{$\pm$}l r@{$\pm$}l r@{$\pm$}l r@{$\pm$}l r@{$\pm$}l r@{$\pm$}l r@{$\pm$}l r@{$\pm$}l r@{$\pm$}l}
    \hline\hline
    Filter                  &   
    \multicolumn{2}{c}{$A$} &   
    \multicolumn{2}{c}{$B$} &   
    \multicolumn{2}{c}{$C$} &   
    \multicolumn{2}{c}{\ainv}  &  
    \multicolumn{2}{c}{$\left( {\rm d}\pr/{\rm d}\alpha\right)\vert_{\alpha = \ainv}$} &   
    \multicolumn{2}{c}{$\alpha_{\rm min}$} &  
    \multicolumn{2}{c}{$P_{\rm min}$}      & 
    \multicolumn{2}{c}{$\alpha_{\rm max}$}&  
    \multicolumn{2}{c}{$P_{\rm max}$}     \\ 
    \hline
    B      & 12.540 & 0.227 & 0.574 & 0.108 & $-0.390$ & 0.044 & 18.8 & 1.3 & 0.12 & 0.01 & 6.9 & 0.4 &$-0.77$& 0.30 & 112.1 & 2.3 & 15.04 & 0.22 \\
    V      & 12.479 & 0.177 & 0.543 & 0.085 & $-0.308$ & 0.038 & 19.8 & 0.8 & 0.12 & 0.01 & 7.0 & 0.5 &$-0.88$& 0.21 & 110.7 & 2.1 & 14.32 & 0.18 \\
    R      & 12.232 & 0.147 & 0.470 & 0.069 & $-0.321$ & 0.031 & 19.6 & 0.7 & 0.13 & 0.01 & 6.3 & 0.4 &$-0.99$& 0.21 & 112.3 & 1.7 & 14.22 & 0.14 \\
    \hline
  \end{tabular}
  Note\,: there are no observations in the negative branch, hence the \amin\ and \pmin\ values are obtained from extrapolation by the trigonometric model in Equation.~(\ref{Eqn:trigo}).
  \end{center}
\end{table*}

A Levenberg-Marquardt algorithm allowed us to recover the best-fit values for the $A$, $B$, $C$, and $\alpha_{\rm 0}$ parameters, as well as the corresponding covariant matrix. This matrix leads to the $1\,\sigma$ uncertainty on the fit parameters. For a given set of fit parameters, the so-called polarimetric slope $h$, defined as the derivative of modelled phase-polarisation curve $\pr(\alpha)$ at the inversion angle \ainv\ can be computed analytically by\,:
\begin{equation}
h = A\sin^B(\ainv)\cos^C\left(\frac{\ainv}{2}\right)
\label{eqn5}
\end{equation}
The positions $(\amin,\amax)$, and values $(\pmin,\pmax)$ of the minimum and maximum of the modelled phase-polarisation curve can be computed by finding numerically the zeroes of the derivative of $\pr(\alpha)$.
To estimate the $1\,\sigma$ uncertainty of these quantities, a Monte-Carlo approach has been used. A number $N=10000$ of quadruplets $(A,B,C,\ainv)$ have been drawn randomly, according to Gaussian distributions with mean and standard deviations given by the values and uncertainty delivered by the Levenberg-Marquardt fit. For each of theses random realizations, the quantities $h$, $\amin$, $\amax$, $\pmin$, and $\pmax$ are calculated. Their mean and standard deviations provide estimates of the values and uncertainties of these quantities. The resulting values are listed in Table~\ref{tab:Coeffs}.
The inversion angle and the slope at inversion angle have very similar values in R, V, and B bands\,: $\ainv\sim 19.5^\circ$ and $h\sim 0.12$\%/deg (see Table~\ref{tab:Coeffs}). Note that no polarimetric data are available in the negative branch of the phase-polarisation curve. Thus, the values of $\ainv$ and $h$ reported here, have been derived through model-fitting. They are thus model-dependent.

The polarimetric slope $h$ is an important parameter characterising a phase-polarisation curve, as it yields an estimate of the geometric albedo $\rho$
through semi-empirical relationships such as\,:
\begin{equation}
\log_{10}\left(\rho\right) = C_1
\log_{10}\left(h\right) + C_2
\label{eqn4}
\end{equation}
where $C_1$ and $C_2$ are band-dependent empiric calibration constants. Several sets of $(C_1,C_2)$ calibration constants can be found in the literature. See \textit{e.g.} \citet{Celetal15}, and a thorough synthesis by \citet{Lupishko2018}.
In the $V$ filter, the empiric constants $C_1 = -1.111 \pm 0.031$ and $C_2 = -1.781 \pm 0.025$ in \citet{Celetal15} would lead to the albedo estimate $\rho_V= 0.175 \pm 0.040$. The constants $C_1 = -1.016 \pm 0.010$ and $C_2 = -1.719 \pm 0.012$ in \citet{Lupishko2018} (weighted mean number 2) would lead to $\rho_V= 0.165 \pm 0.033$. Both estimates are compatible.

It is worth emphasizing that this estimate of the geometric albedo is both calibration-dependent through the values of $C_1$ and $C_2$, and model-dependent through the
value of $h$ which comes out of an extrapolation by the trigonometric model~(\ref{Eqn:trigo}), since no measurements on \FAtt\ are available in the negative branch.

\subsection{Maximum positive polarisation}
Our polarimetric measurements of \FAtt\ allow us to estimate the maximum positive fraction of linear polarisation \pmax, a determination that is possible only in rare cases when the object is observed at phase angles up to at least $100-110^\circ$. This parameter potentially provides an important constraint for the modelling of the asteroid’s surface properties, such as its surface regolith grain size \citep{Ito2018}. Figure~3 of \citet{GeaDol86} shows an empirical relationship between \pmax\ and albedo for different sizes of particles. According to this plot, our \pmax\ $\simeq14$\% and albedo $\simeq0.17$ tells us that the particles are large, at least larger than $200\,\mu$m.

\subsection{Polarisation versus wavelength}
\label{spectropol}
Table~\ref{tab:ObservingLog} shows that the broadband values are generally quite close to each other, with the polarisation in the B filter slightly higher than in the other filters. This situation is called ``blue polarisation''. Figure~\ref{Fig_PolSpectrum} shows the polarisation spectrum obtained at phase angles $\simeq 97^\circ$, normalised by its values in the V filter \citep[such a normalisation makes the spectrum fairly independent on the phase angle at which it was obtained, see][]{Bagetal15}. The spectrum of \FAtt\ is nearly flat, exhibiting a very mild overall concave shape, with a negative wavelength gradient in the spectral regions covered by the B, V, R filters
($\lambda \sim 400 - 700$\,nm), and a positive wavelength gradient at longer wavelengths ($\lambda \sim 750 - 900$\,nm). At its red end, our polarimetric spectrum nearly reaches the same high values as those at its blue end. We note that the full spectrum out to about 0.9\,nm reveals a more complex pattern than what the B, V, R filters alone suggest. This indicates that reliable classification and characterisation of small bodies in the solar system would benefit from spectropolarimetry over a wide wavelength range, ideally extending into the IR.

\subsection{Searching for rotational modulation}
\begin{figure}
\centerline{\includegraphics[angle=0,width=9.0cm,clip]{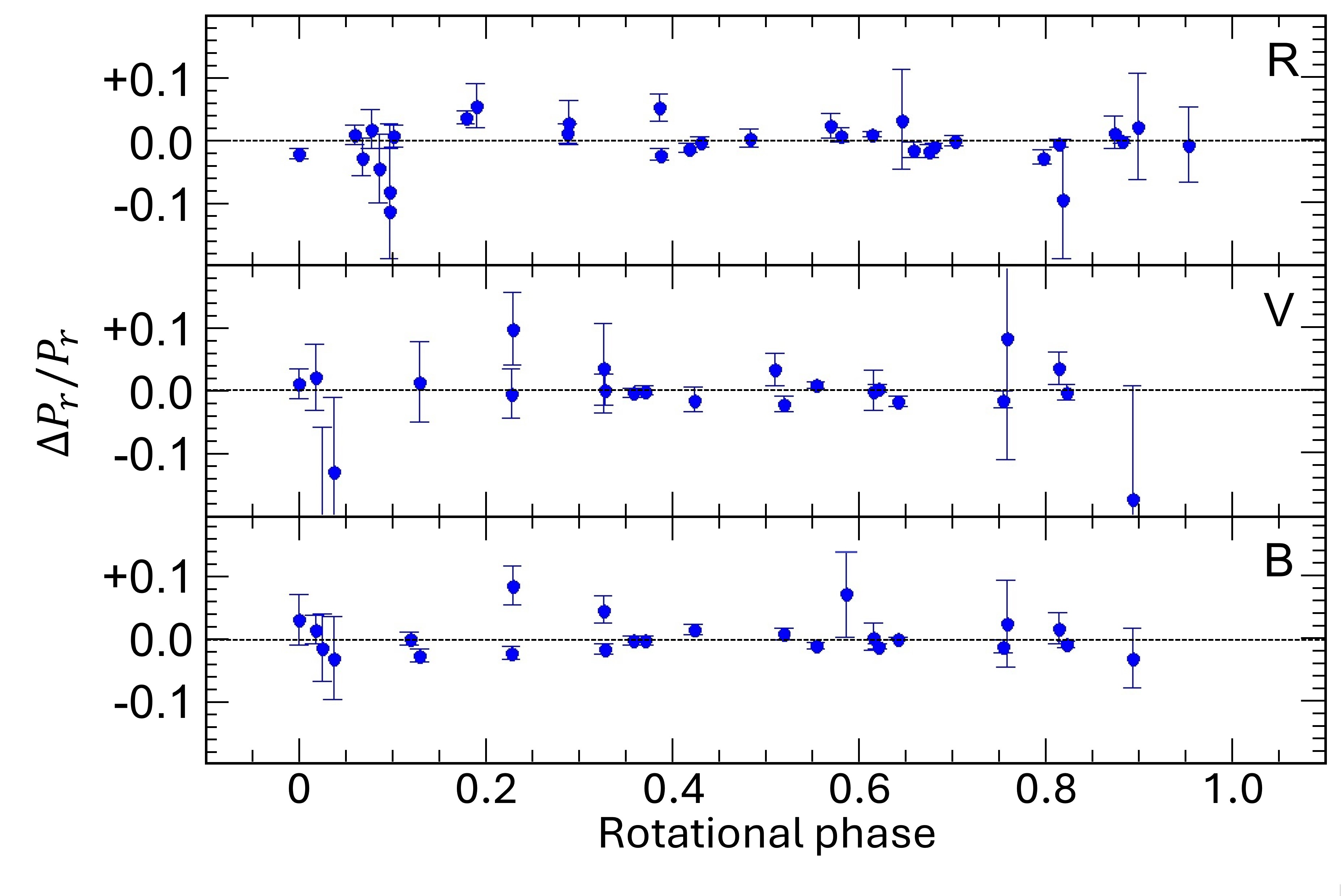}}
\caption{Normalized residual polarisation fraction (residuals between the observed polarisation and the best fit, normalised to the fitted values), plotted as a function of the rotational phase of \FAtt.}
\label{Fig_Rot}
\end{figure}

Rotational modulation of asteroid polarisation is rarely observed, which suggests that surface structure usually does not vary enough to produce detectable changes at small phase angles. A clear exception is Vesta, where ultra-precise measurements revealed such variability \citep{WikNor15,Celetal16}. NEAs offer better chances because they can be observed at large phase angles, where the intrinsic polarisation is higher and potential variations stand out more clearly \citep{Borisov_2018,Devogele_2024}. For \FAtt, however, we found no evidence of rotational modulation.
Figure \ref{Fig_Rot} shows the residuals between the observed polarisation and the best fit, normalised to the fitted values, plotted against rotational phase. We adopted a rotation period of $13.075$~h \citep{Nath26} and set phase zero at 2025 09 18 UT 00:00.

\section{Comparison with other asteroids}\label{Sect_Comparison}

The polarisation spectrum of \FAtt\ resembles the nearly wavelength-independent behaviour shown by some C-class asteroids reported by \citet{Kwonetal23}.
This similarity, however, does not imply a compositional link. The Umov law \citep{Umov05} tells us that polarimetric spectra often track the inverse of reflectance spectra. Within any taxonomic class the range of spectral slopes can be wide. C-type asteroids, for example, span steep, moderate, and nearly flat reflectance slopes \citep{DeMeo2009}. Their inverse trends differ accordingly. Because of this variability, matching \FAtt\ to any specific C-type asteroid would be unreliable. \FAtt\ lacks the low albedo typical of
the C-class, typically well below $0.08$, and its polarisation phase curve does not show the steep behaviour expected for that group.

Figure~\ref{Fig_PolSpectrum} shows a striking resemblance between the polarisation spectrum of \FAtt\ and that of main-belt F-class asteroid (45) Eugenia. However, the inversion angle of the polarisation curve of \FAtt, $\ainv \sim 20^\circ$ is typical of the large majority of the main-belt asteroids. This rules out a link to other classes of low-albedo asteroids, including B and, more markedly, F, which exhibit \ainv\ values below $20^\circ$ \citep{Belskaya2017}. Moreover, a few of these asteroids  reach values of $\pmax \ga 40$\,\% (B-type asteroid 3200 Phaeton; \citealt{Devetal18}), or exceed 20\,\% already at phase angle $\sim 50^\circ$ \citep[F-type asteroid 101955 Bennu;][]{Celetal18}, well above what we measure for \FAtt.  We can also exclude asteroids with unusually large inversion angles, such as the Barbarians, which show $\ainv > 30^\circ$ \citep[e.g.,][]{Celetal06, Celetal18}.

The albedo of \FAtt\ is only marginally consistent with that of several S-class asteroids. These objects, though, show a slope reversal in their polarisation spectra, which occurs at longer wavelengths than in \FAtt. (Their reversal appears near 760\,nm, while \FAtt\ shows a minimum near 700\,nm). The polarimetric slopes of S-class are also much steeper than those measured here, both where the gradient is positive and where it is negative (see, as an example the polarimetric spectrum of the S-class NEA (433) Eros shown in Fig.~\ref{Fig_PolSpectrum}). Furthermore, S-types reach at most $\pmax \simeq 7-8$\,\% \citep{Ishetal17,Kisetal24,Shcetal25}, much lower than the value for \FAtt.

An interesting fact is the resemblance of some polarimetric properties of \FAtt\ with those for the lunar highlands, e.g., they have similar albedo, characteristics of the polarisation minimum, and value of \pmax; also, some young features demonstrate similar wavelength behaviour of polarisation \citep{Shkuratov15}. However, they differ in the inversion angle and location of the maximum.

Although the problem of finding asteroidal analogues of the polarimetric spectrum of \FAtt\ seems to be difficult, We note that there is a clear similarity with the polarimetric spectrum of the large (about 100 km in diameter) asteroid (21) Lutetia, originally a member of the Tholen M class \citep{Tholen84} and later reclassified as Xc in the SMASS system \citep{BusBin02,Demetal09}. Xc asteroids typically show almost flat reflectance and polarisation spectra \citep{Bagetal15}, and some exhibit
$\ainv \sim 20^\circ$ \citep{Canetal12}.

The polarimetric spectrum of Lutetia is displayed as a red line in Fig.~\ref{Fig_PolSpectrum}, and it is easy to see that its spectrum fits quite well the polarisation spectrum of \FAtt. According to current knowledge, (21) Lutetia and \FAtt\ share also comparable albedos, but their phase polarisation curves exhibit some relevant differences. Lutetia inverts near $25^\circ$ \citep{Celetal16}, while \FAtt\ does so around $20^\circ$.

(21) Lutetia’s former taxonomic classification is interesting, because M class asteroids were long thought to be metal rich \citep{Matter2013}. The spectroscopic survey of M types by \citet{Foretal10} shows a wide range of reflectance slopes whose shapes resemble the inverse of \FAtt’s polarisation spectrum. This makes an M (or Xc) class assignment plausible.
\citet{Belskayaetal22} conducted a general investigation of the polarimetric properties of asteroids that had been originally assigned to the M taxonomic class. They found that, when looking at the relationship between the depth of the negative polarisation branch and the inversion angle, they could subdivide M-class asteroids into two main sub-types, possibly related to different surface compositions.
One subgroup seems to be similar to irons and stony-iron
meteorites, while the second subgroup, which includes (21) Lutetia, seems to be more similar to enstatite and iron-rich carbonaceous chondrites.

\section{Conclusions}
After earlier NEA polarimetric observations collected between 2020 and 2023 within the NEOROCKS project,\footnote{see https://www.neorocks.eu/} these new observations of \FAtt\ represent one of the most extensive and systematic polarimetric studies of a NEA at large phase angles. Although they sample only the positive polarisation branch, the data are dense and accurate enough to determine the key polarimetric parameters, including the inversion angle \ainv, the polarimetric slope $h$, and the maximum positive polarisation \pmax, reached at a phase angle of about $110^\circ$. We also obtained the first spectropolarimetric measurements of a small Solar System body near its polarisation maximum. No convincing evidence is found for surface heterogeneity producing measurable polarimetric effects.

These results constrain the surface properties of \FAtt\ and provide a valuable test case for rapid-response polarimetric techniques applied to newly discovered NEAs. 
We derive an albedo of $0.17 \pm 0.04$ in V band, intermediate between typical $C$-class and $S$-complex asteroids. This value, together with the polarisation spectrum, is consistent with some M (or Xc)-type asteroids. Combined with forthcoming photometric measurements of the absolute magnitude of \FAtt, this albedo will allow an estimate of the object’s size.

\begin{acknowledgements}
Based on observations obtained with data collected at the Paranal Observatory under programs ID 115.29F0.002 and 115.29F0.003 (filter polarimetry and spectropolarimetry of \FAtt, PI Ieva), 095.C-0925(A) (spectropolarimetry of Lutetia and Eugenia, PI=Bagnulo) and 092.C-0639(A) (spectropolarimetry of Eros, PI=Cellino).  G.B.\ acknowledges partial support from grant K$\Pi$-06-H88/5 by the Bulgarian National Science Fund.

G.B.\ gratefully acknowledges observing grant support from the Institute of Astronomy and the National Astronomical Observatory, Bulgarian Academy of Sciences.

BNAO Rozhen is a facility of the National Roadmap for Research Infrastructure 2020-2027 (RACIO project), financially coordinated by the Ministry of Education and Science of the Republic of Bulgaria.

S.V.B. acknowledges the support of the ERC Advanced Grant HotMol  ERC-2011-AdG-291659 for the construction of the DIPol-UF polarimeter, in collaboration with the University of Turku, Finland.

This project is funded by the Horizon Europe Programme of the European Union and implemented by ESA. Views and opinion expressed are however those of the authors only and the European Commission cannot be held responsible for any use which may be made of the information contained therein.

\end{acknowledgements}

\bibliography{references}

\appendix
\onecolumn
\section{Observing log of the polarimetric measurements}
\label{App_ObservingLog}
\setlength{\tabcolsep}{0.06cm}
\begin{longtable}{ccrrr@{$\pm$}lr@{$\pm$}lr@{$\pm$}lr@{$\pm$}lr@{$\pm$}lc}
\caption{\label{tab:ObservingLog}
Observing log for the 2025 FA22 polarimetric campaign.
}\\
\hline\hline
Date       & UT    & Exp  & $\alpha$    & \multicolumn{6}{c}{\pr (\%)} &
\multicolumn{2}{c}{$PC$(B,V)} & \multicolumn{2}{c}{$PC$(V,R)} & Instrument \\
YYYY-MM-DD & hh:mm & (s)  & ($^\circ$)  &
\multicolumn{2}{c}{B} & \multicolumn{2}{c}{V} & \multicolumn{2}{c}{R} &
\multicolumn{4}{c}{(\% per nm)} \\
\hline
\endfirsthead

\caption[]{Observing log (continued).}\\
\hline\hline
Date       & UT    & Exp  & $\alpha$    & \multicolumn{6}{c}{\pr (\%)} &
\multicolumn{2}{c}{$PC$(B,V)} & \multicolumn{2}{c}{$PC$(V,R)} & Instrument \\
YYYY-MM-DD & hh:mm & (s)  & ($^\circ$)  &
\multicolumn{2}{c}{B} & \multicolumn{2}{c}{V} & \multicolumn{2}{c}{R} &
\multicolumn{4}{c}{(\% per nm)} \\
\hline
\endhead

\hline
\multicolumn{15}{r}{\emph{Continued on next page}}\\
\endfoot

\hline\hline
\endlastfoot
2025-09-18 & 02:49 &  1440 & 109.43 &  \nullc      & \nullc       & 13.92 & 0.12 &\nullc             &\nullc             & FoReRo2    \\
2025-09-18 & 03:42 &  2880 & 107.47 & 15.46 & 0.60 & 14.02 & 0.35 & 14.33 & 0.22 &$-0.0152$ & 0.0073 & $+0.0031$ & 0.0041& DIPol       \\
2025-09-18 & 05:09 &   440 & 104.24 & 14.88 & 0.16 &\nullc        & 14.59 & 0.14 &\nullc             &\nullc             & ALFOSC      \\
2025-09-18 & 08:16 &   240 &  96.98 & 14.33 & 0.10 & 13.71 & 0.10 & 13.44 & 0.10 &$-0.0065$ & 0.0015 & $-0.0027$ & 0.0014& FORS2 (PMOS)\\
2025-09-18 & 08:26 &   240 &  96.58 & 14.29 & 0.10 & 13.73 & 0.10&  13.54 & 0.10 &$-0.0059$ & 0.0015 & $-0.0019$ & 0.0014& FORS2 (PMOS)\\
2025-09-18 & 23:55 &  2880 &  65.49 &  9.15 & 0.05 &  9.04 & 0.05 &  9.00 & 0.04 &$-0.0012$ & 0.0007 & $-0.0004$ & 0.0006& DIPol       \\
2025-09-19 & 00:47 &  2880 &  64.20 &  8.87 & 0.04 &  8.72 & 0.05 &  8.57 & 0.04 &$-0.0016$ & 0.0007 & $-0.0015$ & 0.0006& DIPol       \\
2025-09-19 & 01:41 &  2880 &  62.89 &  8.69 & 0.10 &  8.27 & 0.07 &  8.37 & 0.07 &$-0.0044$ & 0.0013 & $+0.0010$ & 0.0010& DIPol       \\
2025-09-19 & 02:18 &   720 &  61.95 &  \multicolumn{4}{c}{}       &  7.97 & 0.09 &\nullc             &\nullc             & FoReRo2     \\
2025-09-19 & 02:32 &  2880 &  61.68 &  8.32 & 0.09 &  8.04 & 0.11 &  8.09 & 0.05 &$-0.0029$ & 0.0015 & $-0.0005$ & 0.0012& DIPol       \\ 
2025-09-19 & 03:25 &  2880 &  60.50 &  8.09 & 0.04 &  7.88 & 0.10 &  7.89 & 0.04 &$-0.0022$ & 0.0011 & $+0.0001$ & 0.0011& DIPol      \\
2025-09-19 & 03:38 &   40  &  60.34 & \multicolumn{4}{c}{}        &  8.03 & 0.66 &\nullc             &\nullc             & ALFOSC \\
2025-09-19 & 21:49 &  4320 &  45.27 &  4.63 & 0.05 &  4.56 & 0.18 &  4.70 & 0.07 &$-0.0007$ & 0.0020 & $+0.0014$ & 0.0019& DIPol \\
2025-09-19 & 23:07 &  4320 &  44.58 &  4.51 & 0.04 &  4.44 & 0.11 &  4.40 & 0.04 &$-0.0007$ & 0.0012 & $-0.0004$ & 0.0012& DIPol \\
2025-09-20 & 00:22 &  4320 &  43.93 &  4.51 & 0.04 &  4.23 & 0.08 &  4.38 & 0.06 &$-0.0029$ & 0.0009 & $+0.0015$ & 0.0010& DIPol \\
2025-09-20 & 01:38 &  4320 &  43.30 &  4.34 & 0.04 &  4.07 & 0.05 &  4.27 & 0.05 &$-0.0028$ & 0.0007 & $+0.0020$ & 0.0007& DIPol \\ 
2025-09-20 & 02:39 &  1440 &  42.77 &  \multicolumn{4}{c}{}       &  4.06 & 0.05 &\nullc             &\nullc             & FoReRo2\\
2025-09-20 & 02:53 &  4320 &  42.71 &  4.19 & 0.09 &  4.03 & 0.13 &  4.04 & 0.04 &$-0.0017$ & 0.0017 & $+0.0001$ & 0.0014& DIPol \\ 
2025-09-20 & 05:28 &   240 &  41.85 &  4.06 & 0.10 &  3.99 & 0.10 &  3.98 & 0.10 &$-0.0007$ & 0.0015 & $-0.0001$ & 0.0014& FORS2 (PMOS)\\
2025-09-20 & 21:14 &  6120 &  37.01 &  3.03 & 0.07 &  2.92 & 0.15 &  3.00 & 0.09 &$-0.0012$ & 0.0017 & $+0.0008$ & 0.0017& DIPol \\
2025-09-20 & 21:33 &  1440 &  36.92 &  \nullc      &  \nullc      &  2.95 & 0.05 &\nullc             &\nullc             & FoReRo2 \\
2025-09-20 & 22:43 &  6120 &  36.67 &  2.84 & 0.03 &  2.83 & 0.18 &  3.04 & 0.10 &$-0.0001$ & 0.0019 & $+0.0021$ & 0.0021& DIPol  \\
2025-09-21 & 00:00 &  6120 &  36.37 &  3.10 & 0.09 &  3.00 & 0.16 &  2.90 & 0.10 &$-0.0011$ & 0.0018 & $-0.0010$ & 0.0019& DIPol  \\
2025-09-21 & 01:16 &  6120 &  36.08 &  2.93 & 0.06 &  2.77 & 0.19 &  2.91 & 0.06 &$-0.0017$ & 0.0020 & $+0.0014$ & 0.0019& DIPol  \\
2025-09-22 & 03:40 &    28 &  32.20 & \nullc       & \nullc       &  2.07 & 0.04 &\nullc             &\nullc             & FORS2 (IPOL) \\
2025-09-22 & 03:44 &    40 &  32.19 & \nullc       &  2.00 & 0.05 & \nullc       &\nullc             & $+0.0007$ & 0.0006& FORS2 (IPOL) \\
2025-09-22 & 22:09 &  6120 &  30.31 &  1.75 & 0.12 &  1.73 & 0.31 &  1.52 & 0.16 &$-0.0002$ & 0.0035 & $-0.0021$ & 0.0035& DIPol \\
2025-09-22 & 23:55 &  6120 &  30.16 &  1.63 & 0.08 &  1.30 & 0.28 &  1.64 & 0.10 &$-0.0035$ & 0.0030 & $+0.0034$ & 0.0029& DIPol \\
2025-09-23 & 01:41 &  6120 &  30.00 &  1.63 & 0.09 &  1.19 & 0.26 &  1.55 & 0.09 &$-0.0046$ & 0.0029 & $+0.0036$ & 0.0028& DIPol \\
2025-09-23 & 22:02 &  3240 &  28.58 &  1.51 & 0.10 &  1.59 & 0.27 &  1.42 & 0.11 &$+0.0008$ & 0.0030 & $-0.0017$ & 0.0029& DIPol \\
2025-09-24 & 03:36 &    60 &  28.30 & \nullc       & \nullc       &  1.29 & 0.04 &\nullc             &\nullc             & FORS2 (IPOL) \\
2025-09-24 & 04:14 &  2400 &  28.19 & 1.30 & 0.09  & \nullc       &  1.16 & 0.13 &\nullc             &\nullc             & ALFOSC \\
2025-09-28 & 03:59 &    80 &  24.17 & \nullc       & \nullc       &  0.60 & 0.07 &\nullc         &\nullc                 & FORS2 (IPOL) \\
2025-09-28 & 04:05 &   100 &  24.17 & \nullc       &  0.52 & 0.07 &  \nullc      &\nullc         & $+0.0008$ & 0.0010    & FORS2 (IPOL) \\

\hline 
\end{longtable}

\noindent
FORS2 observations in spectropolarimetric mode (PMOS) are reported giving the polarisation values convolved with the response curves of the filters used in the DIPol-UF instrument. FORS2 (IPOL) refers to VLT measurements obtained with filter polarimetry.

Minor Planet Center (MPC) observatory codes are as follow. DIPol@C2PU (Calern): R87; FORS2@VLT (Cerro Paranal): 309; FoReRo2@2mRCC (Rozhen): 071; ALFOSC@NOT (La Palma): Z23.

\end{document}